\newif\ifpreprint

\preprinttrue 

\ifpreprint
\documentclass[aip,jcp,amsmath,amssymb,preprint]{revtex4-1}
\else
\documentclass[aip,jcp,amsmath,amssymb,reprint]{revtex4-1}
\fi

\usepackage[utf8]{inputenc}
\usepackage{amsmath}
\usepackage{amssymb}
\usepackage{xspace}
\usepackage{graphicx}
\usepackage{dcolumn}
\usepackage{braket}
\usepackage{float}
\usepackage{color}
\usepackage{xcolor}
\newcommand{\code}[1]{{\normalfont\ttfamily #1}\xspace}

\usepackage{hyperref}
\hypersetup{
    colorlinks,
    linkcolor={red!50!black},
    citecolor={red!70!black},
    urlcolor={red!80!black}
}

\usepackage{listings}
\definecolor{codegreen}{rgb}{0.58,0.4,0.2}
\definecolor{codegray}{rgb}{0.5,0.5,0.5}
\definecolor{codepurple}{rgb}{0.25,0.35,0.55}
\definecolor{codeblue}{rgb}{0.30,0.60,0.8}
\definecolor{backcolour}{rgb}{0.98,0.98,0.98}
\definecolor{mygray}{rgb}{0.5,0.5,0.5}

\definecolor{sqred}{rgb}{0.85,0.1,0.1}
\definecolor{sqgreen}{rgb}{0.25,0.65,0.15}
\definecolor{sqorange}{rgb}{0.90,0.50,0.15}
\definecolor{sqblue}{rgb}{0.10,0.3,0.60}

\lstdefinestyle{mystyle}{
    backgroundcolor=\color{backcolour},
    commentstyle=\color{codegreen},
    keywordstyle=\color{codeblue},
    numberstyle=\tiny\color{codegray},
    stringstyle=\color{codepurple},
    basicstyle=\ttfamily\footnotesize,
    breakatwhitespace=false,
    breaklines=true,
    captionpos=b,
    keepspaces=true,
    numbers=left,
    numbersep=5pt,
    numberstyle=\ttfamily\tiny\color{mygray},
    showspaces=false,
    showstringspaces=false,
    showtabs=false,
    tabsize=2
}

\newcommand{\tens}[3]{{#1}_{#2}^{#3}}

\newcommand{\cop}[1]{\hat{a}^\dag_{#1}}
\newcommand{\aop}[1]{\hat{a}^{\phantom \dag}_{#1}}
\newcommand{\sqop}[2]{\hat{a}_{#2}^{#1}}
\newcommand{\aphystei}[2]{\braket{#1 \| #2}}

\newcommand{\no}[1]{ \{ {#1} \}}
\newcommand{\orbspace}[1]{\mathbb{#1}}

\newcommand{\adensity}[2]{\bar{\gamma}_{#2}^{#1}}
\newcommand{\tdensity}[4]{[\boldsymbol\gamma_{#1#2}]_{#4}^{#3}}
\newcommand{\Eh}[0]{$E_\mathrm{h}$\xspace}

\newcommand{\forte}[0]{\textsc{Forte}\xspace}
\newcommand{\psifour}[0]{\textsc{Psi4}\xspace}
\newcommand{\psithree}[0]{\textsc{Psi3}\xspace}
\newcommand{\pyscf}[0]{\textsc{PySCF}\xspace}
\newcommand{\pybind}[0]{\textsc{Pybind11}\xspace}
\newcommand{\numpy}[0]{\textsc{numpy}\xspace}
\newcommand{\ambit}[0]{\textsc{Ambit}\xspace}
\newcommand{\chemps}[0]{\textsc{CheMPS2}\xspace}

\usepackage[capitalize]{cleveref}

\DeclareUnicodeCharacter{2212}{-}
\DeclareUnicodeCharacter{0301}{\'{e}}
\DeclareUnicodeCharacter{0308}{\"}

\begin{document}

\title{Forte: A Suite of Advanced Multireference Quantum Chemistry Methods}

\author{Francesco A. Evangelista}
\email{francesco.evangelista@emory.edu}
\affiliation{Department of Chemistry and Cherry Emerson Center for Scientific Computation, Emory University, Atlanta, GA 30322, USA}
\author{Chenyang Li}
\affiliation{Key Laboratory of Theoretical and Computational Photochemistry, Ministry of Education, College of Chemistry, Beijing Normal University, Beijing 100875, China}
\author{Prakash Verma}
\author{Kevin P. Hannon}
\author{Jeffrey B. Schriber}
\author{Tianyuan Zhang}
\author{Chenxi Cai}
\author{Shuhe Wang}
\author{Nan He}
\author{Nicholas H. Stair}
\author{Meng Huang}
\author{Renke Huang}
\author{Jonathon P. Misiewicz}
\author{Shuhang Li}
\author{Kevin Marin}
\author{Zijun Zhao}
\affiliation{Department of Chemistry and Cherry Emerson Center for Scientific Computation, Emory University, Atlanta, GA 30322, USA}
\author{Lori A. Burns}
\affiliation{Center for Computational Molecular Science and Technology, School of Chemistry and Biochemistry, School of Computational Science and Engineering, Georgia Institute of Technology, Atlanta, Georgia 30332-0400, USA}

\date{\today}

\begin{abstract}
\forte is an open-source library specialized in multireference electronic structure theories for molecular systems and the rapid prototyping of new methods.
This paper gives an overview of the capabilities of \forte, its software architecture, and examples of applications enabled by the methods it implements.
\end{abstract}

\maketitle

\section{Introduction}
\label{sec:introduction}
Software development plays a critical role in advancing quantum chemistry,\cite{Sherrill.2010.10.1063/1.3369628} enabling the exploration of new fundamental theoretical ideas and modeling systems of ever-increasing complexity.\cite{Matthews.2020.10.1063/5.0004837,Folkestad.2020.10.1063/5.0004713,Peng.2020.10.1063/5.0005889,Aquilante.2020.10.1063/5.0004835,Manni.2023.10.1021/acs.jctc.3c00182,Guther.2020.10.1063/5.0005754,Apra.2020.10.1063/5.0004997,Barca.2020.10.1063/5.0005188,Sun.2020.10.1063/5.0006074,Seritan.2020.10.1063/5.0007615,Oliveira.2020.10.1063/5.0012901,Saue.2020.10.1063/5.0004844,Lischka.2020.10.1063/1.5144267,Werner.2020.10.1063/5.0005081,Kallay.2020.10.1063/1.5142048,Neese.2020.10.1063/5.0004608,Balasubramani.2020.10.1063/5.0004635,Smith.2020.10.1063/5.0006002,Nakano.2020.10.1063/5.0005037,Mutlu.2023.10.1063/5.0142433,Chan.2024.10.1063/5.0196638}
In the past decade, the availability of quantum chemistry packages that use modular designs and provide Application Programming Interfaces (APIs) has enabled the creation of specialized software plugins, enhancing the capabilities of the original codes.\cite{Turney.2012.10.1002/wcms.93,Parrish.2017.10.1021/acs.jctc.7b00174,Smith.2018.10.1021/acs.jctc.8b00286,Sun.2018.10.1002/wcms.1340,Sun.2020.10.1063/5.0006074,Lehtola.2023.10.1063/5.0175165,Hicks.2024.10.1063/5.0190834}
The availability of well-documented APIs is particularly beneficial in the context of academic scientific software development because it reduces the entry barrier for new developers and shields them from the complexities of large software projects.

A prime example of these benefits is \forte, an open-source plugin developed in our group for the quantum chemistry package \psifour.\cite{Turney.2012.10.1002/wcms.93,Parrish.2017.10.1021/acs.jctc.7b00174,Smith.2020.10.1063/5.0006002}
Initially a C++ plugin for \psithree,\cite{Crawford.2007.10.1002/jcc.20573} \forte has transformed into a sophisticated software suite with a Python interface, illustrating the evolution and integration facilitated by modern APIs.
The rich quantum chemistry functionality provided by \psifour's API and the continued improvements made by its dedicated developer community have allowed the developers of \forte to concentrate efforts on pioneering new quantum chemistry methods, accelerating its evolution.

Forte specializes in the implementation of multireference electronic structure theories,\cite{Szalay.2012.10.1021/cr200137a,Lyakh.2012.10.1021/cr2001417,Mazziotti.2012.10.1021/cr2000493}  focusing on methods that can be applied to a variety of challenging problems, including predicting global potential energy surfaces and modeling open-shell spin-coupled systems and multi-electron excited states.
Multireference methods assume a correlated reference wave function as a starting point for describing a target electronic state.
This reference state is assumed to account for static electron correlation effects, which can be captured only by a linear combination of Slater determinants.
A major challenge in electronic structure theory is creating multireference approaches that can achieve high accuracy, especially in scenarios where single-reference methods like coupled cluster theory fall short.\cite{Lyakh.2012.10.1021/cr2001417,Kohn.2013.10.1002/wcms.1120,Evangelista.2018.10.1063/1.5039496}
Issues with the numerical stability and generality of the resulting methods have been significant obstacles in virtually all attempts to generate coupled-cluster theory to the multireference case.
Perhaps the most important distinguishing feature of \forte is its implementation of a promising class of multireference methods based on the Driven Similarity Renormalization Group (DSRG). \cite{Li.2019.10.1146/annurev-physchem-042018-052416}
The DSRG methods implemented in \forte mitigate the numerical instabilities connected with the intruder state problem in perturbation theory, are systematically improvable, yield continuous potential energy surfaces, have lower computational requirements than similar methods, and can be scaled to large active spaces.
Besides implementing DSRG methods, \forte also offers a variety of complementary tools, ranging from standard and adaptive active space solvers, multireference quantum embedding methods, orbital transformations, and a convenient framework for the rapid prototyping of new many-body methods.

In this article, we provide a concise review of the theoretical methods underpinning \forte (\cref{sec:methods}), describe its software design principles and capabilities  (\cref{sec:structure}), and showcase representative applications (\cref{sec:examples}). We conclude by discussing current limitations and future development plans (\cref{sec:limitations}).

\section{Methods implemented in \forte}
\label{sec:methods}
An overview of the main methods and functionalities implemented in \forte is reported in \cref{table:methods}.
Forte specializes in implementing multireference quantum chemistry methods based on the active space formalism and its generalizations.\cite{Roos.1980.10.1016/0301-0104(80)80045-0}
The basic assumption of active space methods is that out of the spin orbital basis $\{ \psi_p \}$ (taken to be orthonormal), only a subset of orbitals $\orbspace{A}$ termed ``active'' are essential to approximate the set of electronic states targeted.
The remaining spin orbitals are assumed to play a secondary role and are taken to be either occupied by an electron (core, $\orbspace{C}$) or unoccupied (virtual, $\orbspace{V}$).
Note that this partitioning need not be based on the orbital energy, although in many chemical applications, the active orbitals are selected using an energetic criterion.

\forte is a code for molecular computations, and as such, it can handle general fermionic Hamiltonians with up to two-body interactions of the form
\begin{equation}
\label{eq:hamiltonian}
	\hat{H} = V + \sum_{pq} \tens{h}{p}{q} \cop{p}\aop{q}
	+ \frac{1}{4} \sum_{pqrs} \tens{v}{pq}{rs} \cop{p}\cop{q}\aop{s}\aop{r},
\end{equation}
where $V$ is a scalar term, $\tens{h}{p}{q} = \braket{\psi_p | \hat{h} | \psi_q}$ is a matrix element of the one-electron kinetic plus potential operator ($\hat{h}$), and $\tens{v}{pq}{rs} = \aphystei{pq}{rs} = \braket{\psi_p(1) \psi_q(2)| r_{12}^{-1} | \psi_r(1) \psi_s(2)} - \braket{\psi_p(1) \psi_q(2)| r_{12}^{-1} | \psi_s(1)  \psi_r(2)}$ is an antisymmetrized matrix element of the Coulomb operator.

\subsection{Active space solvers}
\label{sec:methods:active_space_solvers}

Active space solvers are one of the fundamental categories of methods implemented in \forte.
Active space solvers primarily provide a zeroth-order description of multideterminantal electronic states and account for static electron correlation within the active orbitals; however, they may also be used to perform high-accuracy computations on systems with few electrons/orbitals.

\subsubsection{Complete and Occupation Restricted Configuration Interaction}
Most active space solvers implemented by \forte are based on configuration interaction (CI) and expand the eigenstates in a determinant basis $M = \{ \Phi_\mu, \mu = 1, \ldots, d \}$ termed model space.
A generic Slater determinant may be written as a product of creation operators labeled by core and active indices acting on the vacuum state ($\ket{-}$),
\begin{equation}
\label{eq:model_space_det}
	\ket{\Phi_\mu} =  \cop{u} \cop{v} \cdots\prod_{m \in \orbspace{C}} \cop{m} \ket{-},
\end{equation}
where $\cop{u} \cop{v} \cdots$ is one of the $d$ ways of creating a given number of electrons in the active orbitals such that $\Phi_\mu$ belongs to the desired spin and point group irreducible representation of the target electronic state.
In the occupation number representation, each model space determinant is mapped onto a vector representing the occupation number ($n_{u_i} \in \{0,1\}$) of each active spin orbital:
\begin{equation}
	\ket{\Phi_\mu} \rightarrow \ket{\underbrace{n_{u_1} n_{u_2} \cdots }_{N_\orbspace{A}}}.
\end{equation}

A general CI state $\Psi_{\alpha}$ is a linear combination of model space determinants with expansion coefficients $c_{\mu\alpha}$:
\begin{equation}
\ket{\Psi_{\alpha}} = \sum_{\Phi_\mu \in M} c_{\mu\alpha}\ket{\Phi_\mu}.
\end{equation}
\forte implements configuration interaction using the string-based approach of Handy\cite{Handy.1980.10.1016/0009-2614(80)85158-x} and the vectorized Bendazzoli--Evangelisti algorithm to build the $\boldsymbol{\sigma}$ vector.\cite{Bendazzoli.1993.10.1063/1.464087}
In string-based codes, each determinant is expressed as the tensor product of two ``strings'' of alpha and beta occupation numbers ($I_\alpha$/$I_\beta$)
\begin{equation}
	\ket{\Phi_\mu} = \ket{I_\alpha} \otimes \ket{I_\beta}.
\end{equation}
When expressed in a basis of alpha and beta strings, the CI  coefficients may be considered as a matrix $c_{\mu\alpha} \equiv c_{I_\alpha,I_\beta}$ labeled by the indices of the two strings.
The efficient implementation of string-based CI relies on functionality to associate a unique numerical index (without gaps) to each string and precomputed lists of determinants connected by one- and two-electron substitution operators.\cite{Sherrill.1999.10.1016/s0065-3276(08)60532-8}
While efficient graphical approaches for addressing strings and determinants exist,\cite{Olsen.1988.10.1063/1.455063} \forte relies instead on a hash table that maps strings to their address, using the C++ \code{std::unordered\_map} container.

\forte also implements occupation-restricted configuration interaction. This is particularly useful to reduce the size of the configuration interaction space by eliminating ``deadwood''\cite{Ivanic.2001.10.1007/s002140100285} or to study electronic states characterized by specific orbital occupations (e.g., core-excited states).
If the active space is further partitioned into disjoint spaces, $\orbspace{A} = \cup_{k=1}^{n} \orbspace{A}_k$, it is possible to truncate the model space to determinants that satisfy a specific set of occupation restrictions.
In \forte, these restrictions are imposed both at the level of $\alpha$/$\beta$ strings and on the allowed combinations of such strings.
The resulting code is capable of handling restricted active spaces (RAS),\cite{Olsen.1988.10.1063/1.455063} generalized active spaces (GAS),\cite{Olsen.1988.10.1063/1.455063,Fleig.2001.10.1063/1.1349076,Ma.2011.10.1063/1.3611401} and occupation-restriced multiple active spaces (ORMAS).\cite{Ivanic.2003.10.1063/1.1615954} 
For all CI codes, it is possible to enable projection of the CI wave function on a subspace with well-defined spin quantum numbers for the total $\hat{S}^2$ and $\hat{S}_z$ spin operators.
The procedure employed by \forte transforms the CI expansion from the Slater determinant basis into the configuration state function basis,\cite{Olsen.2014.10.1063/1.4884786} and performs the Davidson--Liu diagonalization procedure in this basis. This approach has been recently revisited by others.\cite{Olsen.2014.10.1063/1.4884786,Fales.2020.10.1063/5.0005155}

The string-based CI codes implement algorithms to compute $k$-body reduced density matrices ($k$-RDMs) for all CI methods, where $k$ ranges from one to three, both as spinful or spin-free quantities.
Given a CI state $\Psi_\alpha$, in the spinful case these quantities are denoted by $\boldsymbol{\gamma}^\alpha_k$ and their elements are defined as:
\begin{align}
\label{eq:1rdm}
	(\boldsymbol{\gamma}^\alpha_1)_{u}^{x} & = \braket{\Psi_\alpha|\cop{x}\aop{u} |\Psi_\alpha},\\
	(\boldsymbol{\gamma}^\alpha_2)_{uv}^{xy} & = \braket{\Psi_\alpha|\cop{x} \cop{y} \aop{v} \aop{u} |\Psi_\alpha},\\
	(\boldsymbol{\gamma}^\alpha_3)_{uvw}^{xyz} & = \braket{\Psi_\alpha|\cop{x} \cop{y} \cop{z} \aop{w} \aop{v} \aop{u} |\Psi_\alpha},
\end{align}
where the indices $u,v,w,x,y,z \in \mathbb{A}$ run over the active orbitals.
\forte, also supports the evaluation of the one-body transition density matrix ($\boldsymbol{\gamma}^{\alpha\beta}_1$), defined as
\begin{equation}
	(\boldsymbol{\gamma}^{\alpha\beta}_1)_{u}^{x} = \braket{\Psi_\alpha|\cop{x}\aop{u} |\Psi_\beta} \quad u, x \in \mathbb{A},
\end{equation}
which is used to evaluate transition properties between two states $\Psi_\alpha$ and $\Psi_\beta$ expressed in a common spin orbital basis.

\subsubsection{Determinantal and selected configuration interaction}
\label{sec:methods:sci}

A second set of active solvers is based on general CI expansions and employs a determinant-based approach, whereby the CI expansion is expressed directly in terms of determinants selected from the FCI space.
\forte implements the adaptive configuration interaction (ACI),\cite{Schriber.2016.10.1063/1.4948308} a variant of the CIPSI method,\cite{Huron.1973.10.1063/1.1679199} which aims to identify a model space $M$ such that the variational energy of a CI expansion in this space ($E_M$) approximately matches the CASCI energy to withing a user-provided parameter $\sigma$:
\begin{equation}
	|E_\mathrm{CASCI} - E_M| \approx \sigma.
\end{equation}
In ACI, the model space is identified by iteratively growing a set of reference determinants ($P$) and screening determinants $\Phi_\mu$ in its first-order interacting space ($F$), using an estimate of the energetic contributions of each new determinant $\epsilon(\Phi_\mu)$.
This energy estimate is obtained by diagonalizing a 2 $\times$ 2 Hamiltonian matrix built from the current CI state $\Psi_\alpha$ and the determinant under consideration $\Phi_\mu$.
In the screening process, error control is enforced by excluding determinants with the smallest energy estimates such that the sum of these perturbative corrections approximately equals the energy threshold $\sigma$.
Specifically, at each iteration of the ACI procedure we enforce the following condition on the determinants excluded from selection:
\begin{equation}
	\sum_{\Phi_\mu}^{\text{added}} | \epsilon(\Phi_\mu) | \leq \sigma.
\end{equation}
As customary in selected CI, the ACI energy can be improved by including the estimated energy contribution of those determinants not included in the reference space.
Using ACI, we have been able to push CI computations up to active spaces with as many as 30 electrons in 30 orbitals.\cite{Schriber.2018.10.1021/acs.jctc.8b00877}
Extensions of ACI for computing excited states\cite{Schriber.2017.10.1021/acs.jctc.7b00725} and modeling the ultrafast dynamics of electrons have also been implemented in \forte.\cite{Schriber.2019.10.1063/1.5126945}

\subsection{Orbital optimization and state averaging}
\label{sec:methods:orbopt}
\forte implements a general algorithm for the self-consistent orbital optimization of multideterminantal wave functions.
In the most general form, orbital optimization may be formulated as a minimization problem, where the central quantity to optimize is the weighted average of the energy of a set of states $S = \{\Psi_\alpha, \alpha=1,\ldots,n \}$.
\forte implements a general state averaging procedure that allows the user to specify a set of states $S$ with different electron number, spin/point-group symmetry, and restrictions on the occupation of each space $\mathbb{A}_k$.
The corresponding Lagrangian is given by the average energy plus constraints for orbital orthonormality [$\mathcal{W}(\{ \mathbf{C} \})$], variational CI conditions [$\mathcal{X}(\{ \mathbf{c}^\alpha \},\mathbf{C})$], and CI orthonormality conditions [$\mathcal{Y}(\mathbf{c})$]:
\begin{equation}
\begin{split}
\mathcal{L}(\{ \mathbf{c}^\alpha \},\mathbf{C}) =&  \sum_{\alpha=1}^{n} \omega_\alpha \braket{\Psi_\alpha | \hat{H} | \Psi_\alpha}\\
& + \mathcal{W}(\{ \mathbf{C} \}) + \mathcal{X}(\{ \mathbf{c}^\alpha \},\mathbf{C}) + \mathcal{Y}(\mathbf{c}),
\end{split}
\end{equation}
where $\omega_\alpha \geq 0$ is the weight of state $\Psi_\alpha$ and the weights add up to one.

In \forte, the orbital optimization is performed via the limited-memory Broyden--Fletcher--Goldfarb--Shanno (L-BFGS) algorithm, initialized using the modified diagonal Hessian suggested in Ref.~\citenum{Chaban.1997.10.1007/s002140050241}.
To compute the averaged electronic energy, orbital gradients, and the diagonal Hessian, the inactive and active Fock matrices and the two-electron integrals of 
the form $\braket{px|uy}$ are required.
We obtain these integrals using \psifour's atomic--orbital (AO) version of the Coulomb and exchange (JK) build algorithm, which is also used for building the Fock matrix in standard self-consistent-field procedures implemented in \psifour.
For every $(x,y)$ pair, we first evaluate the matrix $J_{\mu\nu}^{(xy)} = \braket{\mu x | \nu y}$ using the Coulomb (J) builder, followed by a two-index integral transformation via the matrix multiplication
$\braket{px|uy} = \sum_{\mu\nu}^{\rm AO} C_{\mu p} J_{\mu\nu}^{(xy)} C_{\nu u}$.
For MCSCF computations, the number of L-BFGS iterations is limited to 6--8 since strict convergence at each iteration is unnecessary.
Note that the Lagrangian may be expressed in terms of contractions of the one- and two-electron integrals with 1- and 2-RDMs averaged over the states. 
Therefore, due to the decoupling of the CI and orbital coefficients in the optimization, all active solvers that generate RDMs up to rank two may be interfaced with an optimizer to obtain variationally optimized orbitals.

\subsection{Multireference theories of dynamical correlation}
\label{sec:methods:dyncorr}

\begin{figure*}[hbt]
\includegraphics[width=6.25in]{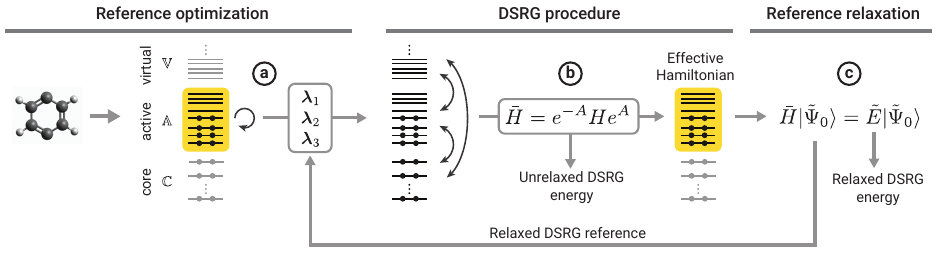}
\caption{Outline of the steps involved in MR-DSRG computations. The first step (a) optimizes the correlated reference state $\Psi_0$ and evaluates its reduced density cumulants ($\boldsymbol{\lambda}_k$, $k$ = 1,2,3).
The cumulants are then passed to the MR-DSRG algorithm (b) to produce the effective Hamiltonian ($\bar{H}$) and its energy expectation value ($\braket{\Psi_0 |\bar{H} | \Psi_0}$). The reference state can be relaxed by diagonalizing $\bar{H}$  to obtain a new reference state $\tilde{\Psi}_0$ (c). The relaxed reference can be used as a starting point for a new MR-DSRG computation, and this procedure may be iterated until self-consistency is reached.}
  \label{fig:dsrg_scheme}
\end{figure*}

\begin{figure}[hbt]
  \includegraphics[width=3.375in]{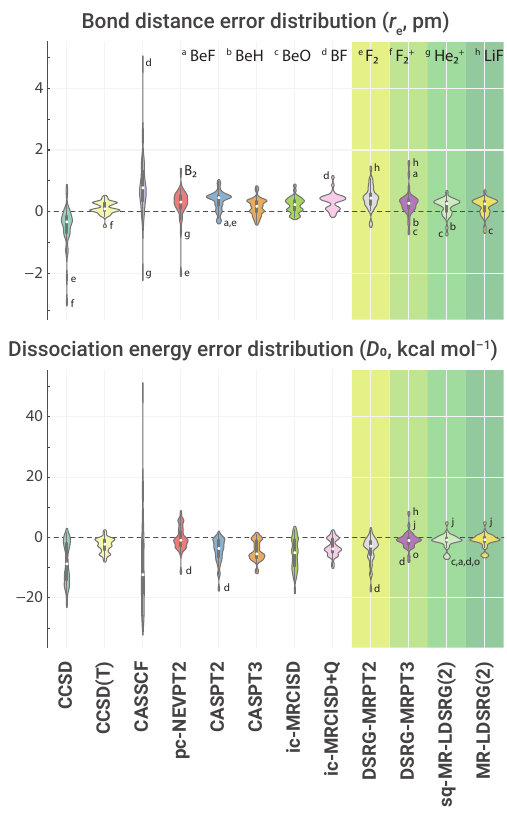}
  \caption{Error distributions for the spectroscopic constants of 33 diatomic molecules. Each violin plot depicts the median (white dot), the interquartile range (thick bar in the center), the upper and lower adjacent values (line in the center), and the probability distribution (width). Molecules with errors lying outside three halves of the interquartile range are labeled. The cc-pCVQZ basis set was employed for Li and Be, while the cc-pVQZ basis set was used for all other atoms.
  Figure adapted from Ref.~\citenum{Li.2021.10.1063/5.0059362}.
  \label{fig:diatomics}}
\end{figure}

Active space solvers provide approximate reference states that do not account for dynamical correlation effects.
A feature unique to \forte is implementing a wide range of dynamical correlation solvers based on the driven similarity renormalization group (DSRG).\cite{Li.2019.10.1146/annurev-physchem-042018-052416}
These solvers can be combined with any active space solver to account for the combined effect of static and dynamic electron correlation.
An overview of the various steps of a multireference DSRG (MR-DSRG) computation is given in \cref{fig:dsrg_scheme}.

The DSRG approach is formally related to unitary multireference coupled cluster theory\cite{Hoffmann.1988.10.1063/1.454125} and analogous approaches.\cite{White.2002.10.1063/1.1508370, Yanai.2006.10.1063/1.2196410, Mazziotti.2006.10.1103/physrevlett.97.143002}
In the DSRG, the second-quantized (bare) Hamiltonian [\cref{eq:hamiltonian}] is brought to a block diagonal form by a unitary transformation that depends on a time-like parameter $s$:
\begin{equation}
	\hat{H} \mapsto \bar{H} = e^{-\hat{A}(s)} \hat{H} e^{\hat{A}(s)}.
\end{equation}
The unitary transformation is written as the exponential of the anti-Hermitian operator $\hat{A}(s)$, preserving the Hermitian character of the bare Hamiltonian in the resulting transformed Hamiltonian $\bar{H}$.
The parameter $s$ controls the extent to which the Hamiltonian is transformed.
For $s > 0$, the DSRG method is designed to gradually suppress those excitations with energy denominators smaller than the energy cutoff $\Lambda = s^{-1/2}$, providing a way to mitigate the intruder state problem.

The anti-Hermitian operator is expressed in terms of an $s$-dependent cluster operator $\hat{T}(s)$ as
\begin{align}
\hat{A}(s) = \hat{T}(s) - \hat{T}^\dag (s),
\end{align}
where $\hat{T}(s)$ is a sum of many-body operators, $\hat{T}(s) = \hat{T}_1(s) + \hat{T}_2(s) + \cdots$, and a generic $k$-body term $\hat{T}_k(s)$ is written in terms of $s$-dependent cluster amplitudes $\tens{t}{ab \cdots}{ij \cdots}(s)$ and normal-ordered second quantized operators ($\no{\sqop{ab \cdots}{ij \cdots}} = \no{\cop{a} \cop{b} \cdots \aop{j}\aop{i}}$) as
\begin{equation}
\label{eq:cluster_operator}
\hat{T}_k(s) =
\frac{1}{(k!)^2} \sum_{ij \cdots} \sum_{ab \cdots} \tens{t}{ab\cdots}{ij \cdots}(s) \no{\sqop{ab \cdots}{ij \cdots}}.
\end{equation}
The cluster amplitudes are antisymmetric with respect to permutations of upper or lower indices, and internal excitations (labeled only by active indices) are implicitly omitted from the definition of $\hat{T}_k(s)$ since they are equivalent to unitary rotations among the model space determinants.

The general goal of the DSRG is to remove those terms in $\bar{H}$ that take determinants in the model space and send them outside of the model space $M$.\cite{Nooijen.1996.10.1063/1.470988,Datta.2011.doi:10.1063/1.3592494,Nooijen.2014.10.1063/1.4866795}
If we express the transformed Hamiltonian in normal-ordered form as:
\begin{equation}
\label{eq:hbar}
\bar{H}(s) = \bar{H}_0(s)
+ \sum_{pq} \tens{\bar{H}}{p}{q}(s) \no{\sqop{p}{q}} + \frac{1}{4} \sum_{pqrs} \tens{\bar{H}}{p q}{r s}(s) \no{\sqop{p q}{r s}} + \ldots,
\end{equation}
then the DSRG aims to remove terms like, for example, $\tens{\bar{H}}{e}{u}(s) \no{\sqop{e}{u}}$ that replace a core orbital $\psi_u$ with a virtual orbital $\psi_e$, transforming a model space determinant [see \cref{eq:model_space_det}] into a new determinant not contained in the model space.
Here normal ordering of the operators is assumed with respect to a correlated state $\Psi_\alpha$ following the approach of Mukherjee,\cite{Mukherjee.1997.10.1016/s0009-2614(97)00714-8} so that the condition $\braket{\Psi_\alpha|\no{\sqop{p q \cdots}{r s \cdots}}|\Psi_\alpha} = 0$ is satisfied for any operator $\sqop{p q \cdots}{r s \cdots}$.
In the DSRG, the amplitudes that enter into $\hat{A}(s)$ are obtained by solving the following set of many-body equations\cite{Evangelista.2014.10.1063/1.4890660}
\begin{equation}
\label{eq:dsrg_flow_mb}
\tens{\bar{H}}{ab \cdots}{ij \cdots} (s) = \tens{r}{ab \cdots}{ij \cdots} (s),
\end{equation}
where $\tens{r}{ab \cdots}{ij \cdots} (s)$ is a function parametrized to match the first-order transformed Hamiltonian elements from the single-reference similarity renormalization group.\cite{Evangelista.2014.10.1063/1.4890660}
After solving the DSRG amplitude equations [\cref{eq:dsrg_flow_mb}], the energy can be evaluated by taking the expectation value of $\bar{H}$ with respect to the reference state $\Psi_\alpha$:
\begin{equation}
	E_\alpha(s) = \braket{\Psi_\alpha |\bar{H}(s)|\Psi_\alpha}.
\end{equation}
Alternatively, it is possible to evaluate a relaxed energy ($E'_\alpha$) and wave function ($\Psi'_\alpha$) by diagonalizing $\bar{H}$ in the basis of model space determinants
\begin{equation}
	\bar{H}(s) \ket{\Psi'_\alpha(s)} = E'_\alpha(s) \ket{\Psi'_\alpha(s)},
\end{equation}
yielding the relaxed reference state
\begin{equation}
	\ket{\Psi'_\alpha(s)} = \sum_{\Phi_\mu \in M} c'_{\mu\alpha} \ket{\Phi_\mu}
\end{equation}
In \forte, the relaxation process may be performed once or alternated to the solution of the amplitude equations until self-consistency is reached.

\forte implements three truncation levels for the DSRG.
Two are perturbative approximations obtained by truncating the correlation energy to second- and third-order perturbation theory (DSRG-MRPT2/3).
These approaches are non-iterative and have a scaling analogous to that of single-reference perturbation theories. In particular, the dominant cost in evaluating the second-order correlation energy is proportional to $N_\orbspace{C}^2 N_\orbspace{V}^2$ (ignoring the cost of transforming the two-electron integrals), while the third-order procedure scales instead as $N_\orbspace{C}^2 N_\orbspace{V}^4$.
For both methods, the term with the steepest scaling with respect to the active space size is of order $N_\orbspace{A}^6 N_\orbspace{V}$.
For these perturbative approaches, a single reference relaxation step can be performed at negligible extra cost, and it is already sufficient to capture the largest improvements to the correlation energy. Given its favorable accuracy/cost ratio, this is the recommended procedure for running perturbative MR-DSRG computations.\cite{Li.2017.10.1063/1.4979016}

A nonperturbative approximation in which the $\hat{A}(s)$ operator is truncated to one- and two-body terms [i.e., $\hat{A}(s) \approx  \hat{A}_1(s) + \hat{A}_2(s)$] is also implemented.
This approach, termed MR-LDSRG(2),\cite{Li.2016.10.1063/1.4947218} truncates each commutator in the Baker--Campbell--Hausdorff (BCH) expansion of $\bar{H}$ to only the zero-, one-, and two-body components:\cite{Yanai.2006.10.1063/1.2196410}
\begin{equation}
\label{eq:lcomm}
[\,\cdot\,, \hat{A}(s)] \approx \sum_{k = 0}^{2} [\,\cdot\,, \hat{A}(s)]_k,
\end{equation}
where $[\,\cdot\,, \hat{A}(s)]_k$ is the $k$-body component of the commutator.
This approximation is then applied recursively to all terms that arise from the BCH expansion.
The cost to evaluate a single commutator [\cref{eq:lcomm}] scales as $O(N_\mathbb{C}^2 N_\mathbb{V}^2 N^2) = O(N_\mathbb{C}^2 N_\mathbb{V}^4 + N_\mathbb{C}^3 N_\mathbb{V}^3 + \ldots)$, with the leading order being identical to the scaling of coupled-cluster with singles and doubles (CCSD) [$O(N_\mathbb{C}^2 N_\mathbb{V}^4)$].
In typical problems, converging the Frobenius norm of $\bar{H}$ to within $10^{-12}$ \Eh requires 10 evaluations of the commutator $[\,\cdot\,, \hat{A}(s)]$, implying that the MR-LDSRG(2) method is about 10 times more expensive than CCSD.
In addition, one must store large intermediate tensors of the size of $O(N^4)$, limiting computation on a single computer node to 200--300 basis functions.
To overcome the limitations of the MR-LDSRG(2) method, the non-interacting virtual orbital (NIVO) approximation is introduced for $[\,\cdot\,, \hat{A}(s)]_2$ in Eq.~\eqref{eq:lcomm}, whereby two-body intermediates labeled by three and four virtual indices are neglected.\cite{Zhang.2019.10.1021/acs.jctc.9b00353}

\forte also implements a sequential version of the MR-LDSRG(2) approximation [sq-MR-LDSRG(2)],\cite{Zhang.2019.10.1021/acs.jctc.9b00353} based on the following stepwise transformation of the Hamiltonian
\begin{equation}
\bar{H}_\mathrm{sq} = e^{-\hat{A}_2(s)} e^{-\hat{A}_1(s)} \hat{H} e^{\hat{A}_1(s)}  e^{\hat{A}_2(s)}.
\end{equation}
The advantage of this approach is that the inner $\hat{A}_1$-dressed Hamiltonian, $e^{-\hat{A}_1(s)} \hat{H} e^{\hat{A}_1(s)}$, can be computed exactly and amounts to a transformation of the one- and two-electron integrals.\cite{Koch.1994.10.1016/0009-2614(94)00898-1,DePrince.2013.10.1021/ct400250u}
When combined with density-fitted or Cholesky decomposed integrals, the cost of this inner transformation scales as $O(M N^3)$, where $M$ is the number of auxiliary basis functions, an effort negligible compared to the cost of evaluating the commutator.
All approximate MR-DSRG schemes implemented in \forte require as input the one- and two-electron integrals and reduced density matrices of the reference state up to rank three, offering an advantage over other multireference methods that require higher-order RDMs.\cite{Zgid.2009.10.1063/1.3132922,Chatterjee.2020.10.1021/acs.jctc.0c00778,Guo.2021.10.1063/5.0051211,Guo.2021.10.1063/5.0051218}
We have also explored the possibility of truncating the three-body RDM in DSRG-MRPT2 to reduce the storage and computational cost further. Our results show that cumulant truncation schemes are particularly useful in applications with large active spaces ($N_\orbspace{A}>$ 10--20).\cite{Li.2023.10.1063/5.0159403}

Extensive benchmarks of the various DSRG methods implemented in \forte have been reported in the literature.\cite{
Li.2016.10.1063/1.4947218,Li.2018.10.1063/1.5023493,Li.2017.10.1063/1.4979016,Li.2018.10.1063/1.5023904,Li.2021.10.1063/5.0059362,Wang.2023.10.1021/acs.jctc.2c00966}
In \cref{fig:diatomics}, we summarize some of the results reported in Ref.~\citenum{Li.2021.10.1063/5.0059362}, which examined error statistics for equilibrium bond distance ($r_\mathrm{e}$) and the dissociation energy ($D_0$, including zero-point vibrational corrections) of 33 open- and closed-shell diatomic molecules computed with spin-adapted MR-DSRG methods.
These computations employ a full-valence CASSCF reference state and the cc-pVQZ and cc-pCVQZ basis sets,\cite{Dunning.1989.10.1063/1.456153,Woon.1995.10.1063/1.470645}
and are compared to coupled-cluster theory with singles and doubles (CCSD),\cite{Purvis.1982.10.1063/1.443164} CCSD with non-iterative triples corrections [CCSD(T)],\cite{Raghavachari.1989.10.1016/s0009-2614(89)87395-6} and various multireference approaches.\cite{Andersson.1990.10.1021/j100377a012,Angeli.2001.10.1063/1.1361246,Werner.1988.10.1063/1.455556}
These results generally indicate that the accuracy of the spin-free MR-DSRG methods follows the order: DSRG-MRPT2 $<$ DSRG-MRPT3 $<$ sq-MR-LDSRG(2) $\lessapprox$ MR-LDSRG(2).
The DSRG-MRPT3, sq-MR-LDSRG(2), and MR-LDSRG(2) methods yield bond distances that are comparable in accuracy with those from CCSD(T), while dissociation energies predicted with the DSRG methods are slightly superior in accuracy.

\subsection{Excited states and spin-adapted implementation}

The DSRG formalism presented thus far corresponds to a state-specific approach with amplitudes optimized for one state at a time.
\forte also implements a state-averaged (or ensemble) generalization of the MR-DSRG that optimizes transformed Hamiltonian for multiple electronic states at a time.\cite{Li.2018.10.1063/1.5019793}
This generalization is useful to 1) describe multiple electronic states and 2) to produce open-shell states of the correct spin symmetry.
In the state-averaged MR-DSRG approach, operator normal ordering with respect to a single state is replaced by a more general normal ordering with respect to a density matrix ($\hat{\rho}$) of all the target states\cite{Kutzelnigg.1997.10.1063/1.474405,Datta.2011.doi:10.1063/1.3592494,Boyn.2021.10.1063/5.0045007}
\begin{equation}
	\hat{\rho} = \sum_\alpha \omega_\alpha \ket{\Psi_\alpha}\bra{\Psi_\alpha}.
\end{equation}
In this generalization for ensembles, a normal ordered operator satisfies the condition $\braket{\no{\hat{O}}}_{\hat{\rho}} \equiv \mathrm{Tr}(\hat{\rho}\no{\hat{O}}) = 0.$
The only practical difference between the state-specific and state-averaged versions of the MR-DSRG is that in the latter, all reduced density matrices are replaced by averaged counterparts:
\begin{equation}
\adensity{pq\cdots}{rs\cdots} = \sum_{\alpha=1}^{n} \omega_{\alpha}
\tdensity{\alpha}{}{pq\cdots}{rs\cdots}.
\end{equation}

When state-averaging is performed on a multiplet of spin states,\cite{Li.2021.10.1063/5.0059362} it is possible to implement spin-free versions of the MR-DSRG equations that have a smaller computational prefactor and reduced memory requirements.
The spin-free MR-DSRG approach produces a transformed Hamiltonian that yields degenerate spin multiplet energies consistent with the state-specific approach and is applicable to states with arbitrary spin quantum numbers.

\begin{table*}[h!]
\caption{Methods and features implemented in \forte.\label{table:methods}}
\begin{ruledtabular}
\begin{tabular}{ll} 
Methods & Notes \\
\hline
FCI & 14--16 active orbitals\\
CASCI/CASSCF & 14--16 active orbitals \\
GASCI/GASSCF & Supports up to 6 spaces\\
ACI/ACISCF & 20--30 active orbitals\\
PCI & 20--100 active orbitals\\
DMRG & Supported via interface to \chemps\cite{Wouters.2014.10.1016/j.cpc.2014.01.019} and Block2\cite{Zhai.2023.10.1063/5.0180424} \\
DSRG methods & All DSRG methods support reference relaxation, spin adaptation, and state-averaging \\
DSRG-MRPT2 & $\sim 2000$ AOs with DF. Analytic energy gradients available for the unrelaxed energy \\
DSRG-MRPT3 & $\sim$750 AOs with DF, $\sim$1450 AOs with DF and FNO \\
LDSRG(2) & $\sim$550 AOs with DF and NIVO, $\sim$1050 AOs with DF, NIVO, and FNO \\
sq-LDSRG(2) & Same as LDSRG(2) \\
ASET(mf) & Environment treated at the mean-field level (CASSCF) \\
Relativistic Hamiltonian & One-electron scalar relativistic Hamiltonian (1e-X2C) supported for all methods via \psifour\cite{Smith.2020.10.1063/5.0006002} \\
AVAS & Supports the selection of atoms, AOs, and planes and average planes for $\pi$ orbitals \\
Integrals & Conventional, density fitting, Cholesky decomposition \\
Frozen natural orbitals & Computed from the DSRG-MRPT2 second-order unrelaxed 1-RDM\\
Orbital localization & Pipek--Mezey supported via an interface to \psifour\cite{Smith.2020.10.1063/5.0006002}
\end{tabular}
\end{ruledtabular}
\end{table*}

\begin{figure*}[hbt]
\includegraphics[width=6in]{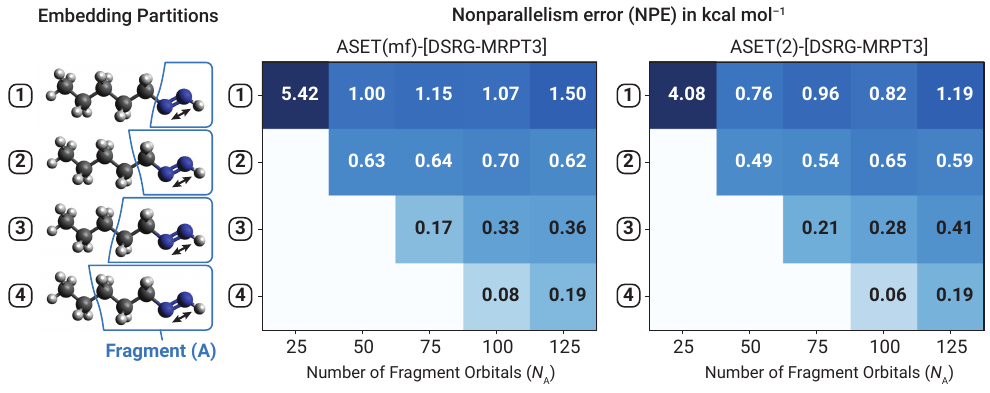}
\caption{Comparison of the ASET(mf) and ASET(2) methods.
Nonparallelism error (NPE) for the pentyldiazene (CH\textsubscript{3}(CH\textsubscript{2})\textsubscript{4}-N=NH) bond dissociation curve computed with DSRG-MRPT3 using different numbers of orbitals in A (NA) and different fragment definitions. All computations here use DSRG flow parameter s = 0.5 $E_\mathrm{h}^{-2}$ and the cc-pVDZ basis.
  Figure adapted from Ref.~\citenum{He.2022.10.1021/acs.jctc.1c01099}.}
  \label{fig:aset_comparison}
\end{figure*}

\subsection{Energy gradients and properties}

Analytic energy gradients with respect to the nuclear coordinates and response properties are implemented for the unrelaxed DSRG-MRPT2 method.\cite{Wang.2021.10.1021/acs.jctc.1c00980}
These can be evaluated either with conventional two-electron integrals or density-fitted ones.
The analytic energy gradients are derived using the method of Lagrange multipliers, where the constraints include the DSRG-MRPT2 energy and amplitude equations augmented with an additional set of conditions that enforce the use of semicanonical CASSCF orbitals.
Because the DSRG-MRPT2 energy is not stationary, neither with respect to orbitals nor the CI coefficients, a linear system of equations that couple orbital and CI coefficients must be solved to account for relaxation effects.
We note that the analytic energy gradients of SA-DSRG-PT2 have been derived by Park\cite{Park.2022.10.1021/acs.jctc.1c01150} and an implementation is available in Bagel.\cite{Shiozaki.2017.10.1002/wcms.1331}

\subsection{Embedding}
\label{sec:methods:embedding}

When dealing with large chemical systems where the properties or chemical changes of interest are spatially localized, computations may be accelerated by embedding a costly high-level computation within a low-level method.\cite{Svensson.1996.10.1021/jp962071j,Richard.2012.10.1063/1.4742816,Manby.2012.10.1021/ct300544e,Isegawa.2013.10.1021/ct300845q,Libisch.2014.10.1021/ar500086h,Sun.2016.10.1021/acs.accounts.6b00356,Wasserman.2020.10.1002/qua.26495}
\forte implements active space embedding theory (ASET),\cite{He.2020.10.1063/1.5142481,He.2022.10.1021/acs.jctc.1c01099} an embedding approach that uses as a starting point an MCSCF state for the entire systems.
In ASET, the system under consideration is partitioned into high- and low-level fragments, here referred to as A and B.
The core and virtual orbitals are then rotated to maximize their overlap with the atomic orbitals centered onto these two fragments, following an approach used in density-matrix embedding theory.\cite{Knizia.2013.10.1021/ct301044e}
In contrast, active orbitals are left unchanged to preserve the energy of the starting MCSCF state.
The partitioning of the orbitals in ASET is controlled by an AO overlap threshold $\tau$ and by the projector $\hat{P}^\mathrm{A} = \sum_{\chi_\mu,\chi_\nu \in \mathrm{A}} \ket{\chi_\mu}(S^\mathrm{A}_{\mu\nu})^{-1}\bra{\chi_\nu}$,
where the basis functions $\chi_\mu$ and $\chi_\nu$ are centered on fragment A and $S_{\mu\nu}^\mathrm{A} = \braket{\chi_\mu|\chi_\nu}$ is the AO overlap matrix.
The overlap of an orbital $\phi'_p$ with the fragment A is measured by the norm of of the projected orbital, that is, $n_p^\mathrm{A} = \braket{\hat{P}^\mathrm{A}\phi'_p| \hat{P}^\mathrm{A}\phi'_p}$.
Rotated orbitals whose value of $n_p^\mathrm{A}$ is greater than $\tau$ are assigned to A, otherwise to B.
Alternatively, the user can select a fixed number of fragment orbitals from those with the largest overlap with fragment A.
\forte implements the lowest level of ASET, termed ASET(mf), which assumes that the total wave function for the entire system ($\Psi_\mathrm{A+B}$) is decomposable in terms of a correlated wave function for A ($\Psi_\mathrm{A}$) and a mean-field state for B ($\Phi_\mathrm{B}$, a Slater determinant):
\begin{equation}
	\Psi_\mathrm{A+B} \approx \Psi_\mathrm{A} \otimes \Phi_\mathrm{B}.
\end{equation}
 This approach treats the interaction between the high- and low-level fragments of the system at the mean-field level, capturing electrostatic and exchange effects.
The ASET(mf) Hamiltonian used in the high-level computation is given by
\begin{equation}
	\hat{G} = E_\mathrm{B} + \sum_{pq}^\mathrm{A} \tens{\tilde{h}}{p}{q} \cop{p}\aop{q}
	+ \frac{1}{4} \sum_{pqrs}^\mathrm{A} \tens{v}{pq}{rs} \cop{p}\cop{a}\aop{s}\aop{r},
\end{equation}
where $E_\mathrm{B}$ is the energy contribution from the occupied orbitals of the low-level part while $\tens{\tilde{h}}{p}{q}$ are effective one-electron integrals that account for the interaction of the electrons in A with the electrons in B, where the latter is the mean-field state.
The generation of the fragment-localized orbitals and the ASET(mf) Hamiltonian corresponds to an orbital transformation and subsequent transformation of the integrals (dropping the occupied and unoccupied orbitals of B).
The ASET(mf) energy is given by the energy of the entire system ($\mathrm{A + B}$) computed at the MCSCF level ($E^{\mathrm{A+B}}_\mathrm{MCSCF}$) plus the correlation energy of A from a high-level multireference computation on the ASET(mf) Hamiltonian ($\delta E^\mathrm{A}_\mathrm{MR}$):
\begin{equation}
	E_\mathrm{ASET(mf)} = E^{\mathrm{A+B}}_\mathrm{MCSCF} + \delta E^\mathrm{A}_\mathrm{MR}.
\end{equation}
Since the cost of ASET(mf) is determined by the projection step, which scales as $N^3$, its cost is negligible compared to that of the MCSCF step.

ASET can be systematically improved, for example, by treating interactions among fragments to second order, like in the ASET(2) method.\cite{He.2022.10.1021/acs.jctc.1c01099} In practice, the increase in accuracy achieved by ASET(2) does not usually justify the extra cost necessary to go beyond a mean-field treatment.
This point is illustrated in \cref{fig:aset_comparison}, where we report the nonparallelism error (NPE)---a measure of the overall deviation of a potential energy surface from a reference one---for the double {N=N} bond dissociation in pentyldiazene, a test case introduced in Ref.~\citenum{Pham.2018.10.1021/acs.jctc.7b01248}.
For this example, the high-level computation uses DSRG-MRPT3 while the embedding Hamiltonian is generated with ASET(mf) and ASET(2).
This plot shows that ASET(2) brings marginal improvements over ASET(mf). These results also suggest that, given a computational budget expressed as the maximum number of fragment orbitals, then the most accurate results are obtained by choosing the largest fragment compatible with such a restriction.

\subsection{Orbital transformations}

\forte implements several types of orbital transformations useful in multireference computations.
These are implemented using a general interface for performing orthogonal transformations of orbitals $\{ \psi_q \}$ into a new set of orbitals $\{ \psi'_p \}$ of the form
\begin{equation}
	\ket{\psi'_p} = \sum_{q} \ket{\psi_q} Q_{qp},
\end{equation}
where $Q$ is an orthogonal matrix ($Q^{-1} = Q^T$).

One type of orbital transformation useful when performing multireference computations is the generation of initial guess orbitals with well-defined valence character.
This transformation can help improve the convergence of MCSCF computations and avoid the problem of multiple solutions.\cite{Meras.1990.10.1016/0009-2614(90)87291-x,Zaitsevskii.1994.10.1016/0009-2614(94)00899-x,Marie.2023.10.1021/acs.jpca.3c00603}
\forte implements the atomic valence active space (AVAS) method,\cite{Sayfutyarova.2017.10.1021/acs.jctc.7b00128} which projects a starting set of orbitals---from Hartree--Fock or density functional theory---onto a set of atomic-like valence orbitals.
This projection is analogous to the one used in ASET, except that the orbitals can be specified by the user via a much more general criterion.
\forte implements a general syntax to let the user specify the target AVAS orbitals. This allows the flexibility to select orbitals using the element symbol, atom number, orbital shell, or even a specific atomic-like orbital.
For example, the following input selects the 3d shell of all Fe atoms and the 2p$_z$ orbital of all C atoms:
\begin{verbatim}
["Fe(3d)","C(2pz)"]
\end{verbatim}
The \forte implementation of AVAS also includes a more recent development in AVAS\cite{Sayfutyarova.2017.10.1021/acs.jctc.7b00128,Sayfutyarova.2024.10.1021/acs.jctc.3c00949} that allows the selection of $\pi$ orbitals by specifying a local orthogonal plane to each local p orbital.

More recently, frozen natural orbitals (FNOs) truncated schemes\cite{Taube.2008.10.1063/1.2902285,Landau.2010.10.1063/1.3276630,DePrince.2013.10.1021/ct400250u} for multireference computations based on second-order perturbation theory were implemented in \forte.\cite{Li.2023.10.48550/arXiv.2312.07008}
This technique has been used to truncate the size of the virtual orbital space, reducing the computational cost of methods that scale steeply with respect to the number of virtual orbitals.
In \forte, we first compute the second-order unrelaxed 1-RDM [\cref{eq:1rdm}] at the DSRG-MRPT2 level and then rotate the virtual orbitals to make the virtual block diagonal.
Once transformed to this basis, the orbitals are sorted according to their natural occupation number, and the virtual frozen orbitals are selected using a cumulative truncation scheme based on cumulative occupation numbers.
The retained virtual orbitals are then transformed to the semicanonical basis before subsequent higher-level computations are performed.
The FNO scheme implemented in \forte even allows the user to specify two different values of the flow parameter, one for the DSRG-MRPT2 and one for the subsequent high-level computation.

\section{Overview of the structure of \forte}
\label{sec:structure}

\begin{figure*}[hbt]
  \includegraphics{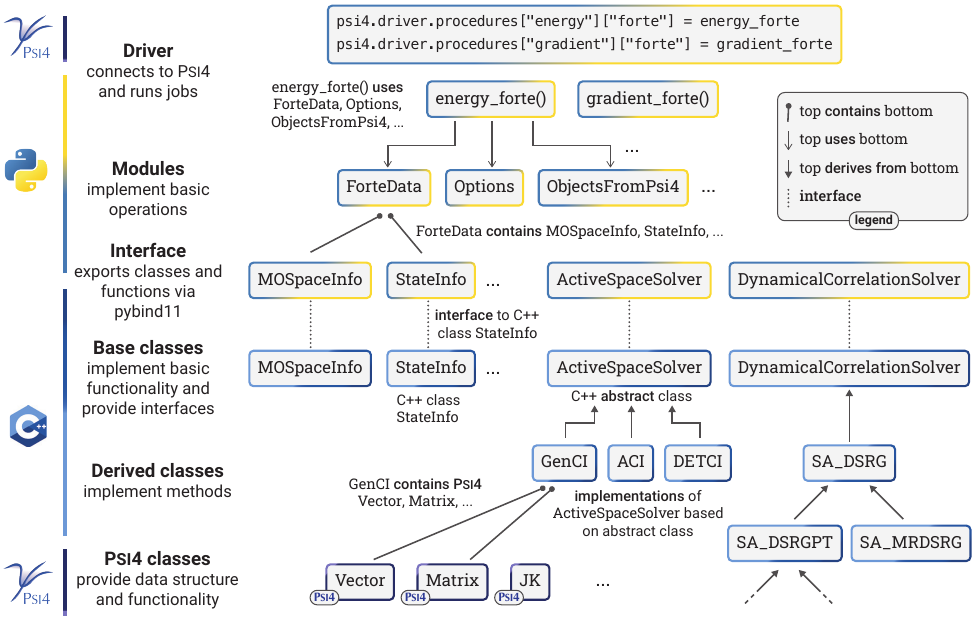}
  \caption{Structure of \forte. The C++ code consists of base classes that provide basic functionality (e.g., managing orbital symmetry) and define interfaces to complex functionality via abstract classes. Derived classes implement active space solvers and dynamical correlation methods. Both base and derived classes contain classes and use functions implemented in \psifour. A \pybind interface exposes the base C++ classes to Python. A higher-level Python layer uses the base classes to implement a modular class system that executes computations. A python driver layer connects \forte to \psifour.  	\label{fig:structure}}
\end{figure*}

An overview of the structure of \forte is given in \cref{fig:structure}.
The main philosophy behind the design of \forte is to create a library of highly-performant functions written using a low-level language, which is exposed to a higher-level language to provide a flexible way to express general computational workflows.
Hence, \forte consists of a collection of C++ classes and methods, a Python interface that exports the C++ functionalities based on \pybind, and Python code that implements several computational workflows using a modular class system.
\forte relies on \psifour in several ways. Many of the C++ methods in \forte use classes and functions provided by \psifour, including vectors, matrices, and JK builders.
The computational workflow of \forte is implemented in Python to leverage \psifour's driver and to delegate many aspects of a computation (e.g., basis set handling, initial generation of mean-field orbitals, geometry optimization).

\subsection{Base classes}

The \forte C++ code is organized around a few base classes used to implement data containers and interfaces for classes that implement computational procedures.
The base classes implemented in \forte are listed in \cref{table:base_classes}.
These consist of classes that store information about the computation and the target electronic states (\code{ForteOptions}, \code{MOSpaceInfo}, \code{StateInfo}, \code{RDMs}) and abstract interfaces for computational methods (\code{ActiveSpaceSolver}, \code{DynamicCorrelationSolver}).

The \code{MOSpaceInfo} class is designed to facilitate the handling of symmetry in \forte, centralizing and simplifying bookkeeping operations.
Throughout \forte, orbitals are assumed to be ordered in the ``Pitzer'' format; that is, orbitals are grouped according to their irreducible representation (irrep) and, within an irrep, are ordered according to their energy.

Classes that implement active space solvers are derived from the abstract class \code{ActiveSpaceMethod}, which has the scope of computing the energy of a manifold of states of the same symmetry (number of electrons, $M_S$, $S$, point group) and restrictions on the occupation of each active space.
The class \code{ActiveSpaceSolver} is then responsible for creating instances of \code{ActiveSpaceMethod} for each different target state.
This allows \forte significant flexibility in generating multiple reference states simultaneously, together with their corresponding averaged reduced density matrices.
This feature enables, for example, state-averaged computations of electronic states of different point group symmetry and spin.
The possibility of averaging over states with different orbital restrictions also enables specialized computations on core-excited states, where occupation restrictions on core orbitals may be used to target specific transitions.\cite{Huang.2022.10.1021/acs.jctc.1c00884,Huang.2023.10.1063/5.0137096}
Similarly, the \code{DynamicCorrelationSolver} class provides a standard interface for dynamical electron correlation methods.
All DSRG methods are derived from this class, and sub-classes are defined to implement different levels of theory.

\begin{table*}[h!]
\footnotesize
\caption{Base classes implemented in \forte.\label{table:base_classes}}
\begin{ruledtabular}
\begin{tabular}{l l} 
 Class name & Description \\
 \hline
 \code{ForteOptions} & User specified options \\ 
 \code{MOSpaceInfo} & Store size and symmetry information of orbitals spaces \\
 \code{StateInfo} & Store information about a target state (electron number, symmetry, spin) \\
 \code{RDMs} & Store reduced density matrices and cumulants \\
 \code{ActiveSpaceSolver} & Manages active space solver computations on multiple target states\\
 \code{ActiveSpaceMethod} & Base class for active space solvers [FCI, GASCI, selected CI, DMRG] \\
 \code{DynamicCorrelationSolver} & Base class for dynamical correlation solvers [DSRG-MRPT2/3, LDSRG(2)]
\end{tabular}
\end{ruledtabular}
\end{table*}

\subsection{Integral classes}

The base class \code{ForteIntegrals} provides a standard interface for computing and accessing integrals in the molecular orbital basis and handles removing doubly occupied and empty orbitals that are excluded from treatments of the correlation energy.
\code{ForteIntegrals} exposes standard one-electron integrals
\begin{equation}
	h_{pq} = \braket{\phi_p | \hat{h} | \phi_q},
\end{equation}
and antisymmetrized two-electron integrals:
\begin{equation}
	\braket{pq\|rs} = \braket{\phi_p \phi_q |\frac{1}{r_{12}} | \phi_r \phi_s} - \braket{\phi_p \phi_q |\frac{1}{r_{12}} | \phi_s \phi_r}.
\end{equation}
The \code{ForteIntegrals} class implements several integral types available from \psifour.
\forte implements conventional two-electron integrals as well as approximate two-electron integrals in the Cholesky\cite{Beebe.1977.10.1002/qua.560120408,Roeggen.1986.10.1016/0009-2614(86)80099-9,Koch.2003.10.1063/1.1578621,Aquilante.2007.10.1063/1.2777146} and density-fitted (DF) form,\cite{Whitten.1973.10.1063/1.1679012,Dunlap.1979.10.1063/1.438313,Vahtras.1993.10.1016/0009-2614(93)89151-7,Feyereisen.1993.10.1016/0009-2614(93)87156-w,Weigend.2002.10.1039/b204199p,Weigend.2009.10.1063/1.3116103} whereby each element is given by the contraction of two three-index tensors ($L_{pr}^P$) over an auxiliary index $P$:
\begin{equation}
	\braket{pq|rs} \approx \sum_P^M L_{pr}^P L_{qs}^P.
\end{equation}
For density-fitted integrals, \forte implements both in-core and disk integral classes, enabling routine applications to systems with 2000--3000 basis functions on a typical single CPU node.
Furthermore, the \code{CustomIntegrals} class handles user-provided integrals that may be read from an external file (e.g., FCIDUMP)\cite{Knowles.1989.10.1016/0010-4655(89)90033-7}, from \pyscf, or can be passed via the Python interface.

In addition, \code{ForteIntegrals} provides methods to evaluate generalized Fock matrices of the form
\begin{equation}
	f_{pq} = h_{pq} + \sum_{rs} \braket{pr\|qs} \gamma^{r}_{s},
\end{equation}
which are used throughout the code by dynamical correlation solvers and orbital optimizers.
The \code{ForteIntegrals} class also automates updating the molecular orbital coefficients (e.g., during orbital optimization) and the corresponding transformation of the one- and two-electron integrals.

\subsection{Support for manipulating determinants, states, and operators} 

The \code{Determinant} class in \forte represents Slater determinants using the occupation number representation and finds use in the CI codes.
This class uses a binary representation of determinant based on array of 64-bit unsigned integers, with each bit representing the occupation number of a spin orbital.
For an array of $k$ 64-bit unsigned integers, this class stores the occupation numbers in the following way
\begin{equation}
	\underbrace{(\underbrace{n_{63}^1\cdots n^1_0}_{\text{64-bit integer}}, \ldots, n_{63}^k\cdots n^k_0)}_\text{array of 64-bit integers} = \underbrace{\ket{n^1_0 \cdots n_{63}^1 \cdots n^k_0 \cdots n_{63}^k}}_\text{element of Fock-space basis},
\end{equation}
where the occupation number of a spin orbital $\psi_p$ is such that $n_p \in \{0,1\}$.
Operations on determinants (getting/setting bits, masking, partial sums of bits, etc.) are implemented using efficient bitwise operations.

\forte also implements functions to manipulate general states and operators, a feature that enables the rapid implementation of new many-body theories.
In \forte, this functionality is supported by two classes: \code{SparseState} and \code{SparseOperator}.
A \code{SparseState} object can represent a general state expressed as a linear combination of determinants that may span the entire Fock space associated with the one-electron basis:
\begin{equation}
	\ket{\Psi} = \sum_{\mu \in \mathcal{S}} c_\mu \ket{\mathbf{n}_\mu},
\end{equation}
where the index $\mu$ labels the elements of a general set $\mathcal{S}$ of occupation number vectors $\ket{\mathbf{n}_\mu}$.
For example, the following superposition of determinants (using a notation that reflects the occupation of a spatial orbital)
\begin{equation}
\ket{\Psi} = \frac{1}{\sqrt{2}}\left( \ket{20} + \ket{00} \right)
\end{equation}
can be produced in \forte via the following Python code:
\begin{lstlisting}[language=Python, caption={},label={lst:sparsestate}]
from forte import SparseState, det
c = 1./ math.sqrt(2.0)
psi = SparseState({ det('20') : c,
                    det('00') : c})	
\end{lstlisting}
This code relies on the utility function \code{det} to create \code{Determinant} objects and insert them in a dictionary passed to the constructor of the \code{SparseState} class.

Similarly, the class \code{SparseOperator} can manipulate general second quantized operators of the form
\begin{equation}
\hat{T} = \sum_\mu t_\mu \hat{\tau}_\mu,
\end{equation}
where $t_\mu$ is a scalar and $\hat{\tau}_\mu$ is a product (string) of second quantized operators.
The symbol $\mu$ denotes a multi-index that collects all the creation and annihilation operators.
Due to efficiency considerations, an operator strings $\hat{\tau}_\mu$ is stored in the following canonical form:
\begin{equation}
	\hat{\tau}_\mu =
\underbrace{\cop{p_1\alpha} \cop{p_2\alpha} \cdots}_\text{$\alpha$ creation}
\underbrace{\cop{q_1\beta} \cop{q_2\beta} \cdots}_\text{$\beta$ creation}
\underbrace{\cdots \aop{s_2\beta} \aop{s_1\beta}}_\text{$\beta$ annihilation}
\underbrace{\cdots \aop{r_2\alpha} \aop{r_1\alpha}}_\text{$\alpha$ annihilation},
\end{equation}
where the indices are assumed to be sorted in ascending order, e.g., $p_{i} < p_{i+1}$, etc.
Internally, \forte represents a general string of second quantized operators in the \code{SQOperatorString} class, which holds two sets of $k$ 64-bit unsigned integers, with each bit representing the presence or absence of the corresponding creation or annihilation operator.
General operators may be easily defined by creating a \code{SparseOperator} object and adding terms, as shown in the example below,
\begin{lstlisting}[language=Python, caption={},label={lst:sparseoperator}]
from forte import SparseOperator
op = SparseOperator()
op.add_term_from_str('[1a+ 2b+ 0b- 0a-]',1.0)
op.add_term_from_str('[2a+ 1b+ 0b- 0a-]',-1.0)
\end{lstlisting}
This code produces the following operator
\begin{equation}
\text{\code{op}} = \cop{1\alpha} \cop{2\beta} \aop{0\beta} \aop{0\alpha}
- \cop{2\alpha} \cop{1\beta} \aop{0\beta} \aop{0\alpha}
\end{equation}

Functionality is also implemented to apply an arbitrary \code{SparseOperator} object $\hat{T}$ to an general \code{SparseState} object $\ket{\Psi}$:
\begin{equation}
\ket{\Psi'} \leftarrow\hat{T} \ket{\Psi}.
\end{equation}
In addition, \forte implements functionality to apply the exponential of $\hat{T}$
\begin{equation}
\ket{\Psi'} \leftarrow e^{\hat{T}} \ket{\Psi},
\end{equation}
via the Taylor series of the exponential.
To support the realization of methods relevant to quantum computing, \forte also implements functionality to evaluate the product of exponentials of single \textit{nilpotent} Fermionic operators (such that $\hat{\tau}_\mu^2 = 0$),
\begin{equation}
\ket{\Psi'} \leftarrow \prod_{\mu} e^{t_\mu \hat{\tau}_\mu} \ket{\Psi} = \prod_{\mu} (1 + t_\mu \hat{\tau}_\mu) \ket{\Psi}.
\end{equation}
When the operator strings entering $\hat{T}$ do not commute ($[\hat{\tau}_\mu, \hat{\tau}_\nu] \neq 0$ for some $\mu$ and $\nu$) different orderings of the exponential factors lead to different operators.
Therefore, \forte uses a specialized class (\code{SparseOperatorList}) to store and manipulate lists of operators $(t_1 \hat{\tau}_1, t_2 \hat{\tau}_2, \ldots)$, preserving the order of the operators and allowing for repetitions.
For each function described above, a corresponding version that applies the anti-Hermitian operator $\hat{T} - \hat{T}^\dagger$ is also available to support the implementation of unitary formalisms.

\subsection{Python Modules}

\begin{figure*}[hbtp]
\includegraphics[width=6.5in]{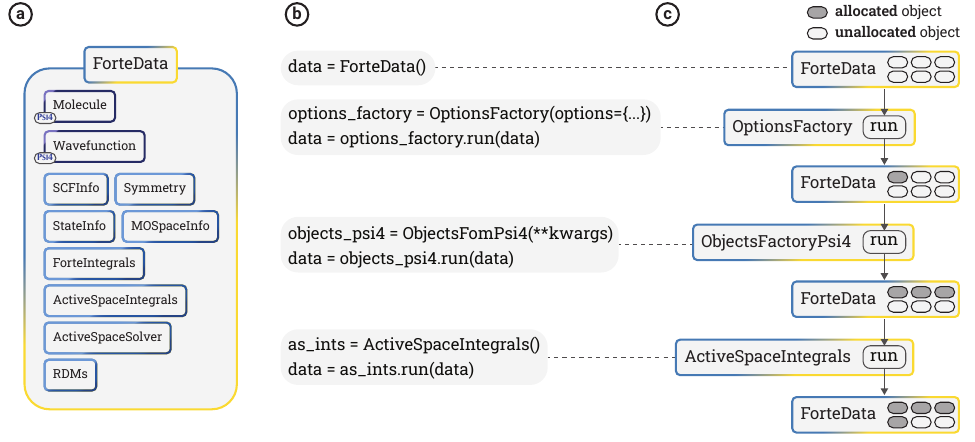}
\caption{The mid-layer Python modules used in \forte. (a) The \code{ForteData} class stores \psifour and \forte base objects that are created and modified during a computation.
(b) Example Python code used in \forte that creates a \code{ForteData} object and uses the \code{OptionsFactory}, \code{ObjectsFromPsi4}, and \code{ActiveSpaceIntegrals} modules. (c) Illustration of the execution flow and how the modules change the \code{ForteData} object from being unallocated (light gray) to allocated (darg gray).}
  \label{fig:modules}
\end{figure*}

The next level of abstraction in \forte consists of a set of mid-level Python modules that provide core functionality, such as setting up \forte objects, performing aspects of a computation, or simplifying access to the Python API.
This layer makes extensive use of the \pybind interface to C++ types and classes and it is easily extended.
Some of the features of this layer are illustrated in \cref{fig:modules}.

Modules define a standard interface for creating steps in a computation and for passing the input and storing the output of a computation.
To standardize their behavior, modules are derived from a base abstract class (\code{Module}).
Modules provide functionality but do not store the input and output data of a computation.
\forte uses the \code{ForteData} class to store all \psifour and \forte objects involved in a computation.
\code{ForteData} is implemented as a Python \code{dataclass}, and after initialization it is empty.

Using modules requires performing two steps:
\begin{enumerate}
	\item \textbf{Module initialization}. In the first step, the module object is created, and parameters are passed to control the behavior of the module:
\begin{verbatim}
module = Module(args)
\end{verbatim}
\item \textbf{Module execution}. In the next step, the member function \code{run} is called passing the \code{ForteData} object
\begin{verbatim}
data = module.run(data)
\end{verbatim}
\end{enumerate}

For example, the module \code{ObjectsFromPsi4} initializes objects used by \forte from \psifour and is invoked in the \forte drivers as
\begin{verbatim}
data = ObjectsFromPsi4(**kwargs).run(data)
\end{verbatim}
The keyword arguments (\code{kwargs}) variable is passed in the constructor of this object to propagate arguments passed by the used in the \psifour function \code{psi4.energy()}.

\subsection{Running \forte: Standard Workflow and Python Interface}

\begin{figure}[hbt]
  \includegraphics{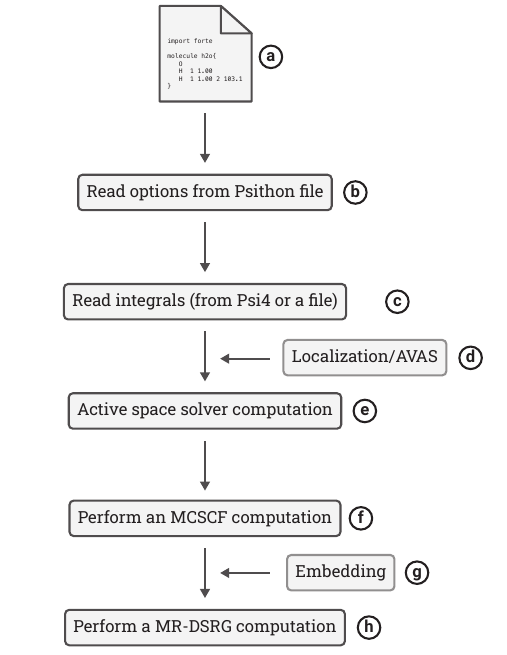}
  \caption{Standard \forte workflow. Starting from an input Psithon file (a), the options controlling the computation are read (b).
In the next step, integrals are read from \psifour or an external file (c).
Optionally, the orbitals are transformed to a localized or AVAS basis (d).
The main computation consists of an initial active space solver computation (e), followed by simultaneous orbital and CI coefficient optimization (f), and a final MR-DSRG computation (h).
If the user requests an ASET(mf) embedding computation, the orbitals are localized prior to performing the MR-DSRG computation (g).
  \label{fig:standard_workflow}}
\end{figure}

The simplest way to run \forte computations is via its interface to \psifour.
This requires the user to write an input file that is then executed by calling \code{psi4} on the command line.
When executed in this way, \forte follows a standard workflow for multireference computations shown in \cref{fig:standard_workflow}.
The main steps in this workflow are an active space solver computation followed by orbital optimization (MCSCF) and an MR-DSRG computation.
The user can optionally transform the initial guess orbitals prior to the MCSCF computation or enable an embedded computation in the MR-DSRG step.
\forte also supports external active space solvers, either through direct integration on the C++ side or via the general \code{ExternalActiveSpaceMethod} class.
This feature has enabled new types of computations, like the ones reported in Ref.~\citenum{Huang.2023.10.1103/prxquantum.4.020313}, where the active space solver was a hardware quantum computation based on the Variational Quantum Eigensolver.\cite{Peruzzo.2014.10.1038/ncomms5213}

This standard workflow encompasses a variety of user case scenarios, allowing any combination of integrals, active space solvers, and dynamical correlation solvers.
This is possible due to \forte's extensive use of abstract interfaces and polymorphism, which enables using a single workflow code that is easy to maintain.
Within this workflow, a computational procedure is expressed in terms of base classes, and the various possible combinations of methods are obtained by instantiating such base classes with derived classes, as specified in the input.

The input for a \forte computation using this standard workflow relies on Psithon, a Python syntax customized for \psifour that simplifies certain aspects of specifying the input to a computation.
 In \cref{lst:dsrg-mrpt2}, we show the input file for a basic state-specific DSRG-MRPT2 computation on the hydrogen fluoride molecule using a CASSCF reference state.
 The main options specified in this input consist of the type of active space (FCI = \code{fci}) and the type of correlation solver (spin-adapted MR-DSRG = \code{sa-mrdsrg}, truncated to second-order = \code{pt2}).
The spin-adapted implementation of the DSRG-MRPT2 is the most computationally efficient and generally recommended for standard computations.
In this example, the $1 a_1$ orbital of HF is optimized in the MCSCF procedure, but it is dropped from the correlated MR-DSRG computation.

\begin{lstlisting}[language=Python, caption={A sample \forte input file for a spin-adapted DSRG-MRPT2 computation on the HF molecule using the cc-pVDZ basis set. Lines 6--11 define the molecular geometry and basis set. Lines 13--21 set the options for \forte. The \code{frozen\_docc}, \code{restricted\_docc}, and \code{active} options specify the number of frozen doubly occupied, the restricted doubly occupied, and active orbitals in each irreducible representation of the $C_\text{2v}$ point group ($A_1,A_2,B_1,B_2$).},label={lst:dsrg-mrpt2}]
# Spin-adapted DSRG-MRPT2 computation on HF using
# a CASSCF(2,2) reference state

import forte

molecule {
  F
  H 1 1.5
}

set basis cc-pvdz

set forte {
  # calculation options
  active_space_solver  fci 		 
  correlation_solver   sa-mrdsrg
  corr_level           pt2
  dsrg_s               1.0
  # orbital subspace dimensions (A1,A2,B1,B2)
  frozen_docc          [1,0,0,0]
  restricted_docc      [1,0,1,1]
  active               [2,0,0,0]
}

energy('forte')
\end{lstlisting}

An alternative way to run \forte is through the Python API directly as shown in \cref{lst:api}.
This example illustrates how the geometry and options may be defined via the \psifour Python API and \forte executed by directly calling the \code{psi4.energy} function.
In the future, we plan to develop code that will enable to run \forte using its Python module infrastructure to enable the user to specify more general workflows than the standard one currently implemented.

\begin{lstlisting}[language=Python, caption={A sample Python script to run a FCI computation using \forte's API. Lines 6--9 define the molecular geometry while Lines 11--17 define the options of this computation. \forte-specific options are prefixed with ``\code{forte\_\_}''. Line 19 runs the standard workflow of \forte.},label={lst:api}]
# FCI computation on Li2 using RHF orbitals

import psi4
import forte

mol = psi4.geometry("""
Li 
Li 1 1.6
""")

psi4.set_options(
    {
        'basis': 'cc-pVDZ',
        'scf_type': 'pk',
        'forte__active_space_solver': 'fci'
    } 
)

psi4.energy('forte',molecule=mol)
\end{lstlisting}

The \forte can also be used through its \pyscf interface as shown in \cref{lst:pyscf}.
This example illustrates how the integrals may be generated via \pyscf and used by \forte by specifying \code{int\_type pyscf} and defining a \code{pyscf\_obj}.

\begin{lstlisting}[language=Python, caption={A sample \forte input file for a spin-integrated DSRG-MRPT2 computation on the HF molecule using its interface for \pyscf. Lines 3--12 run a \pyscf CASSCF computation. Lines 14--24 set the options for \forte. The \code{frozen\_docc}, \code{restricted\_docc}, and \code{active} options specify the number of frozen doubly occupied, the restricted doubly occupied, and active orbitals in each irreducible representation of the $C_\text{2v}$ point group ($A_1,A_2,B_1,B_2$).},label={lst:pyscf}]
import forte
from pyscf import gto, scf, mcscf
molecule ="""
  F
  H 1 1.5
  """
mol = gto.M(atom = molecule, basis = 'cc-pvdz', symmetry = 'c2v')
 	
mf = scf.RHF(mol)
mf.kernel()
mc = mcscf.CASSCF(mf, ncas=2, nelecas=(1,1))
mc.mc2step()
  	
set forte {
  # calculation options
  int_type             pyscf
  active_space_solver  fci 		 
  correlation_solver   dsrg-mrpt2
  dsrg_s               1.0
  # orbital subspace dimensions (A1,A2,B1,B2)
  frozen_docc          [1,0,0,0]
  restricted_docc      [1,0,1,1]
  active               [2,0,0,0]
}

energy('forte', pyscf_obj = mc)

\end{lstlisting}

\subsection{Distribution, Documentation, Tutorials, and External Tools}
Prepackaged binaries for \forte may be obtained from \url{conda-forge.org}, a community-driven repository of conda packages.
\forte is an open-source project, and its codes are distributed for free via GitHub (\url{https://github.com/evangelistalab/forte}) through an LGPL-3.0 license.
Compilation of \forte requires \psifour (\url{https://github.com/psi4/psi4}) and the tensor library \ambit (\url{https://github.com/jturney/ambit}).

Compilation instructions for \forte are available at the GitHub site, while general documentation is available online through the \code{ReadtheDocs} website (\url{https://forte.readthedocs.io}).
Supplementing the \forte documentation is a suite of more than 250 test cases for the methods implemented that demonstrate how to run various types of computations and the relevant user options.
The \forte repository includes a set of tutorials that introduce users to some of the features of the software.
These tutorials cover basic aspects (e.g., running \forte via Jupyter, implementing CCSD using the sparse operator infrastructure), useful functionality (e.g., orbital visualization), and how to implement methods using quantities computed in \forte.

In parallel to the development of \forte, we have created the python package \code{fortecubeview} (\url{https://github.com/evangelistalab/fortecubeview}, installable via \code{pip}), which may be used to display molecular orbitals and normal modes computed with \forte and \psifour.
Another related development is the \textsc{QForte} package,\cite{Stair.2022.10.1021/acs.jctc.1c01155} which implements a quantum computer emulator and library of quantum algorithms and can consume integrals generated by \forte.

\section{Example applications of \forte}
\label{sec:examples}

\subsection{Modeling molecules in complex chemical environments}

\begin{figure}[hbt]
\includegraphics[width=3.375in]{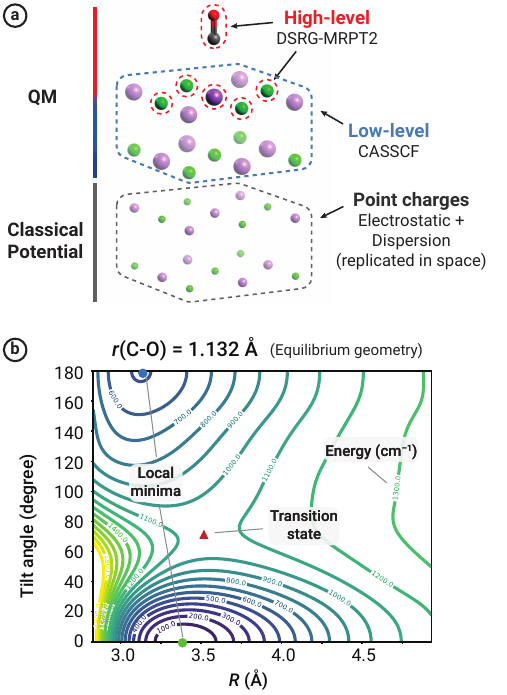}
\caption{(a) Model for the CO-NaCl(100) system generated with the ASET(mf) and DSRG-MRPT2 implementations in \forte.
The quantum mechanical region (QM) is partitioned into a high-level region  {[CO-NaCl\textsubscript{4}]\textsuperscript{3\textminus}} treated at the DSRG-MRPT2 level and a low-level region treated at the CASSCF level, using ASET(mf) embedding.
The CASSCF computation is performed in the presence of a classical point-charge potential (only partially shown) that accounts for electrostatic and dispersion interactions. Carbon and oxygen atoms are represented in gray and red, while {Na\textsuperscript{+}} and {Cl\textsuperscript{\textminus}} ions are colored in purple and green.
(b) Potential energy as a function of the distance of the CO center of mass from the surface ($R$) and the CO tilt angle.
Figure adapted from Ref.~\citenum{He.2023.10.1021/acs.jpca.2c05844}.}
  \label{fig:aset}
\end{figure}

An example of how the tools implemented in \forte may be combined to enable new types of computations is given by our recent multireference study of physisorbed molecules.\cite{He.2023.10.1021/acs.jpca.2c05844}
This work focused on the recent experimental observation of the isomerization of vibrationally excited CO molecules on a NaCl(100) surface.\cite{Lau.2020.10.1126/science.aaz3407} 
To accurately describe the highly vibrationally excited states of CO, we performed multireference computations that can correctly describe the ground state of CO from the equilibrium to the dissociated limit.
As shown in \cref{fig:aset}, we modeled this system by computing the potential energy surface of a single CO molecule on NaCl(100).
The starting point of this computation was a CASSCF computation on the cluster {[CO-Na\textsubscript{9}Cl\textsubscript{9}]} using AVAS guess orbitals that span the full valence space of CO.
This CASSCF computation included a classical external potential to account for electrostatic interactions between the cluster and the {Na\textsuperscript{+}} and {Cl\textsuperscript{\textminus}} ions outside the cluster.
ASET(mf) was then used to compute dynamical correlation energy corrections due to the fragment {[CO-NaCl\textsubscript{4}]\textsuperscript{3\textminus}} at the DSRG-MRPT2 level.
Classical additive corrections were also included to account for dispersion interactions between CO and {Na\textsuperscript{+}} and {Cl\textsuperscript{\textminus}}.
The initial CASSCF step was the main bottleneck in these computations. To map the potential energy surface of CO-NaCl(100), we performed 5376 single-point computations using 288 CPU cores, which ran in less than 50 hours.

\subsection{Simulating X-ray Absorption Spectra}
\begin{figure}[hbt]
\includegraphics[width=3.375in]{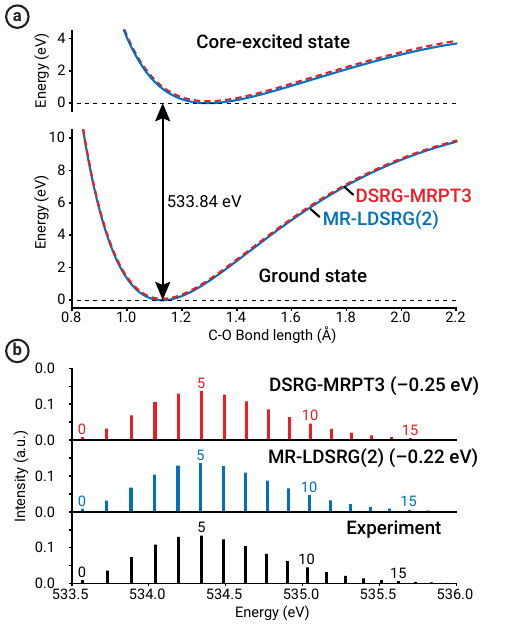}
\caption{(a) Potential energy surface for the ground state and 1s $\rightarrow$ $\pi^*$ core-excited state of CO computed using GAS-DSRG and the cc-pCVQZ-DK basis. All the curves are shifted so that the values of MR-LDSRG(2) potential energy in both traces at each minimum are zero.
(b) Spectrum of CO simulated from the GAS-DSRG potentials and the experimentally determined Morse constants. The theoretical spectra are shifted to align the 0−0 transition with experiment. Each transition is labeled with the vibrational quantum number $\nu'$ of the upper state.
Figure adapted from Ref.~\citenum{Huang.2022.10.1021/acs.jctc.1c00884}.}
  \label{fig:xray}
\end{figure}

Another example of application enabled by \forte is simulating X-ray absorption spectroscopy through our GAS-DSRG scheme.\cite{Huang.2022.10.1021/acs.jctc.1c00884}
This approach starts by describing core-excited states using GASSCF reference states, which capture the dominant orbital relaxation effects.
Following this step, a multi-reference DSRG computation is used to account for dynamical electron correlation.
An illustrative example of the potential application of this approach is our recent computation of the vibrationally-resolved O K-edge 1s $\rightarrow$ $\pi^*$ transition of CO.
In Figure \cref{fig:xray}, we show the simulated spectrum with Franck-Condon factors calculated using one-dimensional potential energy surfaces of both ground and core-excited states obtained from GAS-DSRG and an experimentally-determined Morse potential.\cite{Puttner.1999.10.1103/physreva.59.3415}
Our multi-reference computations accurately describe core-excited CO molecules up to the dissociation limit, as evidenced by the excellent agreement in the vibrational features of CO in the X-ray absorption spectrum and the small energy shifts (less than 0.25 eV) required to align the origin of the spectrum to the experimental value.

\subsection{Rapid prototyping of many-body methods with Forte}

Our second example application of \forte focuses on rapidly prototyping many-body theories.
The \code{SparseState} and \code{SparseOperator} objects allow the implementation of virtually any form of many-body theory expressible in second quantization.
To demonstrate this point, we show how these classes may be used to create a pilot implementation of \textit{arbitrary-order} traditional and unitary coupled-cluster theory using a projective formalism.
At the core of these methods is the evaluation of the residual vector ($r_\mu$):
\begin{equation}
\label{eq:reseq}
r_\mu = \braket{\Phi | \hat{\tau}_\mu^\dagger e^{-\hat{S}} \hat{H} e^{\hat{S}} | \Phi} = 0,
\end{equation}
where the $\hat{S}$ operator is defined as:
\begin{equation}
\hat{S} =
\sum_\mu t_\mu \begin{cases}
 \hat{\tau}_\mu & \text{traditional} \\
 (\hat{\tau}_\mu - \hat{\tau}_\mu^\dagger) & \text{unitary}
\end{cases},
\end{equation}
and the operators $\hat{\tau}_\mu$ that enter $\hat{S}$ are particle-hole excitations with respect to a reference determinant $\Phi$.
Listing~\ref{lst:arbcc} shows that with less than 25 lines of code, \forte can evaluate \cref{eq:reseq} using the functionality provided by the \code{SparseState} and \code{SparseOperator} objects.

\begin{lstlisting}[language=Python, caption={},label={lst:arbcc}]
def cc_residual(op, ref, H, is_antiherm):
	"""Evaluate the CC residual equations"""
    exp = forte.SparseExp()
    
    if is_antiherm:
	    exp_S = exp.apply_antiherm
    else:
    	exp_S = exp.apply_op

    # compute exp(S)|ref>
    wfn = exp_S(op, ref)

    # compute H exp(S)|ref>
    Hwfn = H.apply(wfn)

    # compute exp(-S) H exp(S)|ref>
    R = exp_S(op, Hwfn, scaling_factor=-1.0)

    # compute <q|exp(-S) H exp(S)|ref>
    residual = forte.get_projection(op, ref, R)

    # compute <ref|exp(-S) H exp(S)|ref>
    energy = 0.0
    for det, coeff in ref.items():
        energy += coeff * R[det]
    return (residual, energy)
\end{lstlisting}

A more complex example application of this functionality can be found in Ref.~\citenum{Stair.2021.10.1103/prxquantum.2.030301} where benchmark Projective Quantum Eigensolver computations on {H\textsubscript{10}} systems (corresponding to 20 qubits) were made possible by \forte's sparse state and operator classes.

\section{Current limitations and future development plans}
\label{sec:limitations}

This paper summarizes major developments in \forte over the past decade, highlighting its extensive implementation of DSRG-based methods.
In the future, we plan to improve several aspects of \forte and add code to add desirable functionality currently lacking in the program.
In regards to the functionality supported by \forte, several areas deserve improvement, including the implementation of gradients and static properties beyond state-specific DSRG methods and the implementation of nonadiabatic couplings.
Another set of desirable features includes DSRG-based methods for determining a manifold of excited states.
Our group is currently exploring how to formulate equation-of-motion methods based on the DSRG, a preliminary step required to guide the development of production-level implementations of these new theories.
Another set of functionalities under exploration are 2- and 4-component relativistic extensions of the DSRG methods.
While conceptually straightforward under the no-pair approximation, implementing these methods would require generalizing the code to handle complex tensors.

At the software level, a major plan is to make \forte more interoperable with other quantum chemistry packages to leverage functionality not available in \psifour.
This development will be facilitated by \forte's Python layer, which allows interfacing it with other Python-based packages and writing adaptors to read the information required to perform a multireference computation (e.g., integrals and MO coefficients).
Our group is also working on increasing interoperability, including creating a lightweight library for tensor storage that supports allocating and exporting tensors across C++ and Python without copying data and that supports both real and complex types.
This effort aims to enable the use of existing optimized numerical libraries like \numpy for tensor operations.\cite{Harris.2020.10.1038/s41586-020-2649-2}
Future work will fully integrate this new functionality in \forte.

\section{Conclusion}

By leveraging the power of open-source quantum chemistry codes, the past ten years have seen \forte develop from a simple plugin to \psithree to a suite of advanced multireference electronic structure theories that have pushed the boundaries of quantum chemistry.
We hope that \forte will become a valuable asset to the quantum chemistry community, enabling researchers to perform advanced multireference computations and prototype new theories.
Further collaborations and contributions from the community will be essential in refining \forte's functionalities and expanding its reach to address open challenges in quantum chemistry.

\section*{Data availability}
Data sharing is not applicable to this article as no new data were created or analyzed in this study. 

\begin{acknowledgements}	
The developers of \forte are thankful for the various funding sources that have supported its development. These include the U.S. Department of Energy (under Awards No. DE-SC0016004, DE-SC0019374, DE-SC0024532), the U.S. National Science Foundation (under Awards No. CHEM-1900532, CHEM-2038019, CHEM-2312105), a Research Fellowship of the Alfred P. Sloan Foundation, a Camille Dreyfus Teacher-Scholar Award (TC-18-045), and start-up funds provided by Emory University.
\end{acknowledgements}	

\bibliography{forte.bib}

\end{document}